\documentstyle[12pt,aaspp4]{article}

\begin{document}

\title{Neutron Stars and Black Holes as MACHOs}

\author{Aparna Venkatesan, Angela V. Olinto, and James W. Truran}
\affil{Department of Astronomy and Astrophysics, and Enrico
Fermi Institute, \\ University of Chicago, 5640 S. Ellis
Ave, Chicago, IL 60637}

\authoremail{aparna@oddjob.uchicago.edu,
olinto@oddjob.uchicago.edu, and truran@nova.uchicago.edu}

\begin{abstract}

We consider the contribution of neutron stars and black holes to
the dynamical mass of galactic halos.  In particular, we show
that if these compact objects were produced by an early
generation of stars with initial metallicity $\la 10^{-4}
Z_\odot$, they can contribute at most 30--40\% of the Galactic
halo mass without creating supersolar levels of enrichment. We
show that the case for halo neutron stars and black holes cannot
be rejected on metal overproduction arguments alone, due to the
critical factor of the choice of progenitor metallicity in
determining the yields. We show that this scenario satisfies
observational constraints, similar to but no more severe than
those faced by halo white dwarfs. We also discuss the recent
results on halo microlensing, the presence of enriched hot gas
in clusters and groups of galaxies, and other observations. If
there are halo neutron stars and black holes, they will be
detected by microlensing experiments in the future as
longer-timescale events.

\end{abstract}

\section{Introduction}

Studies of the rotation curves of spiral galaxies and the
dynamical behaviour of galaxy clusters have shown that most of
the matter in our Universe is invisible.  Although most of the
dark matter is believed to be non-baryonic, a significant
component must be baryonic (see, e.g., Copi, Schramm \& Turner
1995). One particular class of baryonic dark matter candidates
consists of objects from about planetary masses to several solar
masses, collectively called MACHOs (massive compact halo
objects). In the last few years, several experiments have used
the gravitational microlensing of stars in the Magellanic Clouds
and in our Galaxy to constrain the nature and amount of such
MACHOs in the Galactic halo.

These experiments have had surprising results. Brown dwarfs,
which were natural candidates for subluminous MACHOs, do not
form a significant fraction of the Galactic halo. In fact,
MACHOs in the mass range 10$^{-7}$ -- 0.02 M$_\odot$ make up
less than 20\% of our halo's dark matter (Renault et
al. 1997). Star counts from the Hubble Deep Field (HDF) along
with microlensing results are consistent with negligible halo
dark matter contributions from substellar objects such as brown
dwarfs (see, e.g., Chabrier \& M\'{e}ra 1997). Furthermore, the
MACHO collaboration observed the microlensing of stars in the
direction of the Large Magellanic Cloud (LMC) by objects which,
based on their two-year data of eight microlensing events, have
a most probable mass of about 0.5$^{+0.3}_{-0.2}$ M$_{\odot}$
(Alcock et al. 1997).  If these objects are in our halo, they
could comprise about half of the halo's mass, based on a
comparison with the optical depth associated with an all-MACHO
halo.

Thus, there are now at least two Galactic dark matter problems:
the nature of halo MACHOs which are detected through
microlensing, and the composition of the remainder of the dark
halo.  The nature of the observed MACHOs is not yet
resolved. There are competing interpretations for the
microlensing data, ranging from non-baryonic halo objects such
as primordial black holes to stellar contributions from a
Galactic component such as the thick disk or spheroid, or from
the warping and flaring of the Galactic disk towards the LMC
(Bennett 1998, and references therein). Other models make a case
for non-Galactic lenses, such as microlensing by stars in the
LMC or SMC themselves (self-lensing), or an intervening
component between the Galaxy and the LMC, such as a dwarf galaxy
or a tidal tail from the LMC. At present, no model gives a
compelling explanation for all aspects of the MACHO problem.

In this paper, we consider baryonic objects as one possible
solution for the MACHO dark matter problem. The most likely
experimental MACHO mass at present is about 0.5 M$_\odot$, which
is too high for typical brown dwarf masses as hydrogen burning
starts at $\sim$ 0.08 M$_\odot$.  0.5 M$_\odot$ suggests the
presence of white dwarfs (WDs) (Adams \& Laughlin 1996; Fields,
Mathews \& Schramm 1997). Alternative candidates for baryonic
MACHOs are neutron stars (NSs), and low-mass black holes
(LMBHs).

Models of WD-dominated halos experience several difficulties.
WD progenitors (taken to be of typical mass 2--4 M$_{\odot}$ in
previous work) produce large amounts of enriched gas, which must
be ejected by some mechanism into the intragroup medium as such
gas is not seen in the Galaxy today. Since neither WD
progenitors nor planetary nebulae have sufficient kinetic energy
in their winds to expel the enriched gas, the presence of
supernovae (SNe) from the death of massive stars (which produce
NSs and BHs) is necessary.  In addition, the lack of the
signature expected from the early luminous phase of WD
progenitors in deep galaxy surveys constrains the WD fraction in
galaxy halos today to be less than 10\% (Charlot \& Silk 1995).
Further constraints on a WD component to the Galactic halo come
from searches for faint stellar objects in the Hubble Deep
Field, down to $I$ = 26.3 (Flynn, Gould \& Bahcall 1996), and
from studying the luminosity evolution and cooling history of
WDs (Graff, Laughlin \& Freese 1998).

Furthermore, given that roughly half of Population I stars are
in binaries, a significant fraction of a WD-dominated halo could
plausibly be in binary systems. These would generate, through
WD-WD mergers, a larger rate of Type Ia SNe than is observed
(Smecker \& Wyse 1991). The chemical enrichment in $^{56}$Fe
from Type Ia SNe would reach far above solar values since the
timescales for Type Ia SNe exceed greatly the initial period of
enhanced star formation, which prevents prompt mixing and
dilution of the ejecta on large scales. These SNe also do not
leave remnants that could lock up the created heavy elements
permanently (Smecker \& Wyse 1991).

NSs and BHs from Type II SNe avoid these difficulties. Type II
SNe occur over much shorter timescales, than do Type Ia, and
they leave a significant fraction of their synthesized heavy
elements in the NS or BH remnants.  In addition, the abundance
ratios of the hot X-ray emitting gas seen in clusters of
galaxies (Mushotzky et al. 1996) appear to favor a Type II,
rather than Type Ia, SN origin, though this is subject to some
debate (e.g., Ishimaru \& Arimoto 1997). Lastly, theoretical
work (Woosley \& Timmes 1996) indicates that an early generation
of massive stars in the Galaxy could indeed have generated a
large number of LMBHs through Type II and Ib SNe explosions.

Models with halo NSs and LMBHs face at least two challenges. The
most likely MACHO mass as reported by Alcock et al. (1997) is
lower than the typical masses of NSs ($\sim$ 1.4 M$_\odot$) and
LMBHs (1.5 to several M$_\odot$).  We address this point in the
discussion section, but note here that the first year's data
from the EROS2 (Exp\'{e}rience de Recherche d'Objets Sombres)
experiment (Palanque-Delabrouille et al. 1998) may indicate
higher MACHO masses.  Secondly, the massive stars that could
create NSs and BHs eject large amounts of enriched material when
they become SNe, which may contaminate the interstellar medium
with $^4$He and heavy elements beyond solar values.

In this paper, we focus on the latter challenge, and show that
the case for a halo with significant mass fractions in NSs and
BHs is not easily dismissed from arguments based solely on
excessive pollution from the ejecta of the SNe creating these
objects. This conclusion is contrary to previous work (see,
e.g. Ryu, Olive \& Silk 1990), primarily because we use recent
results from Woosley \& Weaver (1995; henceforth WW95) for the
nucleosynthetic yields from the evolution of massive stars of
low metallicity that explode as Type II SNe.  These stars do not
have the prohibitively high metal yields that are associated
with solar metallicity progenitors.  Our goal here is to show
that this is therefore not {\it prima facie} evidence to reject
halo NSs and BHs, and to place an upper limit on their
contribution, given the greatest viable stretch of parameters.
Though this scenario is not free of assumptions (\S\ 2), we show
that this upper limit is consistent with other related
constraints.

In order to maximize the contribution of NSs and LMBHs to the
halo dark matter, we focus on an initial burst involving only
high-mass stars and calculate the yields from just this epoch.
This early generation of massive stars create a large population
of compact objects in the initial phase of the formation and
evolution of the Galaxy.  The generated remnants are mostly NSs
and LMBHs, with some heavier black holes. The fraction of NSs
relative to BHs depends on the maximum mass of NSs; LMBHs could
have masses as low as 1.5 M$_\odot$ (Brown \& Bethe 1994).

The plan of this paper is as follows: In \S\ 2, we discuss in
detail the assumptions that we make for a scenario that creates
halo NSs and BHs. In \S\ 3, we compute the yields in hydrogen,
$^4$He, and metals generated by the SNe from this early
population. We derive a maximum mass fraction in the halo made
of remnants from these stars, for three cases of initial
metallicity of the progenitor. We then calculate the mass of gas
required initially for the Local Group to dilute the final
abundances of $^4$He or metals in the leftover gas to solar
values. In \S\ 4, we discuss some implications of this scenario,
relevant constraints, and some recent observations. We then
conclude in \S\ 5.

\section{A Possible NS/BH Scenario}

A halo with a significant fraction of its mass in stellar
remnants must meet certain constraints of mass and
metallicity. The stars that create NSs and BHs also leave behind
ejecta whose enrichment is above solar values, and whose mass
exceeds that in the remnants and therefore those of the present
Galactic disk and bulge. A scenario that proposes NSs and BHs as
significant contributors to halo dark matter must necessarily
invoke at least the following assumptions. A starburst that
selectively forms only massive stars, and in which most of the
baryons participate, is required. Furthermore, this burst must
occur sufficiently early in the Universe's history, as such a
bright phase is not seen at lower redshifts. The prompt
recycling of stellar debris that occurs in standard chemical
evolution is not desirable for a NS/BH scenario, as stars in
present galaxies show a large spread in metallicities and
presumably reflect enrichment from generations subsequent to any
proposed early starburst. Therefore, some mechanism is needed
that removes the initially metal-rich ejecta and that could aid
any dilution that ensues, such as winds or outflows from the
star-forming regions. We discuss these chosen assumptions below;
we will see in \S3 that low-$Z$ stars alleviate any dilution
constraint, compared with high-$Z$ stars.

We assume a dynamical mass of $10^{12}$ M$_\odot$ for a halo of
radius 100 kpc (see, e.g., Peebles 1995, plus the assumption
that $\rho_{\rm halo} \sim r^{-2}$).  Microlensing experiments
constrain the optical depth along the line of sight in MACHOs
from a lensing event, which gives a direct estimate of the total
mass in MACHOs. To extend this to a halo baryonic fraction, we
need to divide this by the total mass of the halo, which is
uncertain and dependent on specific models and assumptions. In
the results of the MACHO experiment, according to Alcock et
al. (1997) and Griest (1997), the halo mass fraction of MACHOs
can take values between 10\% and 100\% for different halo
models, excluding experimental error; however, the most likely
total mass in halo MACHOs within 50 kpc has been very stable at
about 2 $\times 10^{11}$ M$_\odot$. Keeping these considerations
in mind, we derive first the maximum masses in halo MACHOs in
NSs and LMBHs, which we can extend to their halo mass fraction.

In this work, we concentrate on the Local Group (LG) when
deriving limits from specific constraints.  We assume that the
two most luminous, dominant spiral galaxies in the LG, the Milky
Way and M31, each processed $\sim 10^{12}$ M$_\odot$ of
primordial gas through an early generation of massive
stars. This first generation was made of very low metallicity
($\ll Z_\odot$) gas, which cools inefficiently and could favor
the formation of higher stellar masses (Adams \& Laughlin
1997). Furthermore, the maximum stellar mass that is stable
increases with decreasing metallicity. One can approximate this
maximum mass as (Adams \& Laughlin 1997): $ M_{*, max} \approx
114$ M$_\odot \; (1 - 2.4 Z)^2$, which is approximately 114
M$_\odot$ for very small $Z$. From these arguments, we take a
range of progenitor masses 10--100 M$_\odot$, which create NSs
and BHs through Type II SNe. We consider only stars of masses
above 10 M$_\odot$ to maximize the number of NSs and LMBHs, and
assume a Salpeter initial mass function (IMF), where the number
of stars born per unit mass interval, $\xi(M) \propto
M^{-2.35}$.  Although the present-day disk IMF is biased towards
low-mass stars, there is evidence for significant high-mass star
activity in the past.  One possible explanation is a bimodal
star formation history, involving early IMFs skewed towards
high-mass stars (Elbaz, Arnaud \& Vangioni-Flam 1995; Fields,
Mathews \& Schramm 1997). Elbaz et al. (1995) point out that
observations of starburst galaxies, in which the star formation
activity and the relative number of massive stars appear
related, suggest a truncated IMF.

As we show in this work, the composition of the progenitor star
strongly determines that of its ejecta, particularly the
ejecta's metal content. The lower the metallicity of a star, the
weaker is its pollution of the surrounding gas when it becomes a
SN, and the higher the mass of the remnant. Thus, a $Z=0$ star
is the most favorable for creating galactic halos with a
significant fraction in NS/BH remnants. However, self-enrichment
of star-forming regions leads eventually to stars of higher
metallicities. This problem is less serious if the timescale for
the early massive star formation is short compared to the mixing
timescale for the star-forming regions. We will therefore assume
a burst approximation, i.e., there is no immediate recycling of
stellar ejecta as in standard chemical evolution.  The redshift
at which such a burst occurred can be constrained by background
light limits measured today; this is derived in \S4.

The star formation characteristic of the present epoch, where
the rates of SNe are much lower and whose ejecta are mostly
incorporated into new generations of stars, begins later in the
Universe's history. The Galactic disk and halo stars are not
formed in our treatment explicitly; we assume that the baryons
that form such Pop. I and II stars are not enriched by the
earlier starburst, or Pop. III, phase. Such a segregation could
occur if galactic winds accompanying the burst phase removed
associated polluted ejecta, leaving a fraction of material
behind that eventually goes into Pop. I/II stars. As Pop. II
stars have a mean metallicity of about $1/30$ $Z_\odot$, this
segregation of the material forming baryonic MACHOs from that
forming disk and halo stars must be efficient, and is essential
to the scenario discussed here.

In contrast to the case of Pop. I/II stars, the hot X-ray gas
that is seen in many clusters and groups of galaxies contains a
significant fraction, if not most, of the baryons in these
systems, and is enriched in metals to about $Z_X \sim 0.3
Z_\odot$, with some cases approaching solar values (see,
e.g. Mulchaey et al. 1996, Mushotzky et al. 1996). In this work,
we will use this general constraint from studies of this X-ray
gas, rather than the range of observed abundances in Pop. I and
II stars, as we are interested in the remnants from a stellar
phase that involved the bulk of the baryons. We take the degree
of enrichment in intracluster and intragroup gas in other
systems to be roughly solar in metals. As the enriched X-ray gas
exists in systems that also have galaxies containing low-$Z$
stars, a fair fraction of the baryons in such systems may well
have undergone stellar processing separately from the galaxies'
present stars, similar to what we have postulated here. We
discuss in \S 4 how the self-enrichment of disks can be distinct
from any early starburst phase. Note further that the combined
baryonic mass in the Galactic disk and halo stars is at most a
few percent of the total baryons associated with the Galaxy in
its burst phase, and will not alter the main conclusions of this
paper.

In the next section, we use the results of WW95 for the
integrated yield from the early generations of massive stars. We
assume that the SNe associated with this early stellar phase can
effectively drive winds. Two outcomes are possible: the winds
could remove the metal-rich ejecta from the protogalactic
regions and then from the LG itself, or they could mix the
polluted ejecta with the ambient primordial gas outside early
star-forming regions. The second option has the potential to
constrain the mass of baryons that participate in an early
burst, and the final mass in remnants, given a dilution
criterion. We do this for the LG in \S3, where we derive the
baryonic mass required to form the LG in order to avoid
supersolar $^4$He or metal values in the diluted gas remaining
after the first epoch of stellar processing. This provides a
reasonable upper limit to the contribution of halo NSs and BHs
in the extreme case of retaining all the enriched ejecta, i.e.,
no mass loss. We choose solar values as an average final LG
metallicity for the reasons discussed above; though $^4$He
abundances are not measured in the X-ray gas, we include this
case as we will see that $^4$He is more constraining than
metals, as far as dilution requirements are concerned for
low-$Z$ massive stars' ejecta.  Note that if mass loss from the
LG does occur and the enriched undiluted gas is ejected through
galactic winds and outflows, abundances in the leftover gas can
more easily remain at subsolar values. We do not consider this
explicitly in this work, as this would necessitate further
assumptions on the nature, mass and metallicity of the
intergalactic medium.

Finally, we consider the {\it total} mass in NSs and BHs. The
actual fraction of NSs versus BHs depends on the uncertain
maximum mass of NSs. Some authors have recently argued that
stars of masses in the range 18--30 M$_\odot$ could leave
remnant LMBHs after exploding as Type II SNe (Brown \& Bethe
1994; Woosley \& Timmes 1996).  LMBHs may be formed when the
mass of a remnant compact object exceeds 1.5 M$_\odot$ (Brown \&
Bethe 1994). Also, observations may indicate an upper limit to
the NS mass of 1.5 M$_\odot$ (Brown, Weingartner \& Wijers
1995). Finally, simulations of Type II SNe (WW95; Timmes,
Woosley \& Weaver 1996) of low-metallicity stars ($Z \leq
10^{-4} Z_\odot$) with masses above $\sim$ 15 M$_\odot$ show
that their remnants have masses that exceed 1.5 M$_\odot$.  We
therefore derive a minimum mass fraction in NSs by assuming that
the smallest progenitor mass for a LMBH is about 15 M$_\odot$.
If NSs can have masses above $\sim$ 1.5--2 M$_\odot$, the halo
remnants in our results would have fewer BHs.

\section{Results}

In this section, we calculate the IMF-integrated yields for a
burst involving only massive stars of very low $Z$, as described
in \S2, and for $Z = Z_\odot$ stars for comparison. We show that
the enrichment increases with the progenitor's metallicity, with
$Z = Z_\odot$ stars resulting in the familiar extreme metal
pollution, and $Z = 0$ stars having the least enriched ejecta
and the highest mass left in remnants. We then derive the
baryonic requirement for the LG, given the scenario described in
\S2, for dilution of the ejecta from such an early phase to
solar values in $^4$He or metals. Interestingly, the dilution
requirement for $^4$He is more stringent than that for metals.

We use WW95 to calculate the yields for the range of high--mass
stars (10 $\leq$ M/M$_\odot$ $\leq$ 100) of metallicities $Z =
0$ (case 1), and $Z = 10^{-4} Z_\odot$ (case 2). We then fit the
values for the ejected masses in hydrogen, $^4$He, and metals,
as well as remnant masses, as a function of the mass of the star
in the range 10--40 M$_\odot$. We expect minimal mass loss
effects during the main-sequence lifetimes of the
low--metallicity progenitor stars, but there are significant
uncertainites in the energy, location and specific type of
mechanism in launching the shock that leads to SNe in stars more
massive than 40 M$_\odot$. The remnants in these cases are most
likely higher-mass black holes (WW95). The role of fallback and
late-time accretion makes estimating the final remnant mass and
the element yield less reliable. Here we extend the mass
fractions in the four quantities of interest from their values
at 40 M$_\odot$ to the range 40--60 M$_\odot$. Above 60
M$_\odot$, we assume that the entire star collapses into the
remnant.

In Figures 1 and 2, we show the yields of hydrogen, $^4$He,
metals and remnant masses for stars of masses 10--40 M$_\odot$
in cases 1 and 2 respectively, with various symbols and the
solid lines representing the values from WW95 and a polynomial
fit. Figure 1 shows a polynomial fit for the hydrogen and $^4$He
yields only.  For progenitor masses above 25 or 30 M$_\odot$, a
star in the tables of WW95 had more than one case of the final
kinetic energy of the ejecta at infinity, with models A, B, and
C having progressively larger SN shock energies. We chose the
model for which this value was closest to 1.2 $\times$ 10$^{51}$
ergs, i.e., model A in WW95, which had the lowest total ejected
masses compared to models B or C. Clearly, the model with
minimal ejecta will produce remnants with larger masses than one
with maximal ejecta. Figures 1 and 2 show the remnant masses for
the models with maximal ejecta as well, to illustrate the point
that depending on the explosion energy, solar-mass compact
objects can result from stars as massive as 40 M$_\odot$.

From the figures, we see that there is generally more
variability in the yields of zero--metallicity stars (except for
hydrogen), due to their poorly understood structure and
evolution. We obtained good fits only for the hydrogen and
$^4$He yields in case 1, and show the ejected mass in metals and
remnant masses in Figure 1 to emphasize their fluctuations. To
compute metal yields in case 1 integrated over the whole IMF, we
took average values for mass bins of 5 M$_\odot$ over 10--40
M$_\odot$, including separately the peak at 22 M$_\odot$; the
mass left in remnants for case 1 was taken to be the difference
between 10$^{12}$ M$_\odot$ and the integrated post-supernova
output in hydrogen, $^4$He and metals. Above this mass range, we
followed the procedure described earlier.

The $Z = 0$ stars have lower integrated yields in heavy elements
and can create galactic halos with higher mass fractions in NSs
and BHs.  Table 1 shows the computed elemental yields and
remnant masses in solar mass units for cases 1 and 2, resulting
from a stellar population of total mass $10^{12}$
M$_{\odot}$. This is the mass processed initially by each galaxy
in the Group into massive stars. Also shown are the mass
fractions of $^4$He and metals ($Y$ and $Z$) in the ejected gas
relative to solar values, and the fraction of the total halo
mass in remnants, $f \equiv M_{\rm rem}/10^{12}$ M$_\odot$. $f$
should be viewed as an upper limit, as the stellar remnants
which are created here are assumed to eventually reside in the
galactic halos that form later.  For comparison, we include the
case of solar metallicity progenitors (case 3) for both the
minimal ejecta (mE) and maximal ejecta (ME) models from WW95.
We assume $Z_\odot$ = 1.89 $\times$ 10$^{-2}$ and $Y_\odot$ =
0.275 by mass fractions.

From these results, we see that there is more total mass in
remnants, and ejected gas that is less metal-rich, for $Z = 0$
stars as compared to the $Z = 10^{-4} Z_\odot$ stars. For all
cases, the minimum fraction of these remnants in NSs is 24\%,
from stars of mass 10--15 M$_\odot$. Though each set of
progenitors shown in Table 1 has very different metal enrichment
in the ejecta, $Y/Y_\odot$ is roughly the same for all of them.

If we require that the Galaxy and M31 have on average solar
abundances or lower in $^4$He and metals, we need to dilute the
enriched ejecta from each galaxy (Table 1) with a reservoir of
unprocessed gas (at the same metallicity as the progenitors for
cases 1 and 2).  This is done in Table 2 for two values of the
primordial $^4$He abundance: 0.232 (Copi et al. 1995), and the
high value of 0.249 (Tytler, Fan \& Burles 1996).  The initial
mass of gas needed for dilution (including the 2 $\times$
10$^{12}$ M$_\odot$ for the stellar processing in M31 and the
Galaxy) will depend on whether we have the final $Y_{\rm gas}$
or $Z_{\rm gas}$ approach solar values.  Table 2 shows the
baryonic requirement for the entire LG for either purpose for
cases 1 and 2, and the final enrichment in both quantities
relative to solar values. The overall range of initial gas
masses from which the LG formed is 2.4--8.1 $\times 10^{12}$
M$_\odot$.

An alternative for lowering final abundances is to invoke the
loss of most of the metal-rich ejecta from the Group with time,
as is done for some WD models (see, e.g., Fields et al. 1997).
But we see that even without the loss of metal-rich gas, a
significant Galactic halo fraction in NSs and BHs can be
attained if, on average, the gas in the Group has solar
metallicity. Table 2 also shows that, except for one listed
case, the requirement that $^4$He in the leftover gas be at most
solar guarantees metallicities below solar. In other words,
$^4$He is {\it more} constraining than metals, as far as
dilution requirements are concerned for massive stars' ejecta.

Table 2 illustrates the extreme metal enrichment from massive
stars of solar metallicity, and how unfavorable such stars are
for creating a large population of compact objects. For example,
to dilute the metal-rich ejecta from case 3 stars to solar
$Z$-values, the LG must form from primordial gas of initial mass
$\ga 10^{13}$ M$_\odot$. This is prohibitively large, compared
with the corresponding 2.4 $\times 10^{12}$ M$_\odot$ for case
1.

Within the uncertainties in stellar structure and evolution, the
case for a halo with significant mass fractions in NSs and BHs
is not easily dismissed from arguments based solely on excessive
contamination of the intragroup medium with heavy elements. Such
arguments pose serious difficulties for higher metallicity
progenitor stars such as the $Z = Z_\odot$ case. In contrast,
NSs and BHs produced by very low metallicity massive stars can
exist in the Galactic halo, without overproducing metals or
$^4$He. There are however several other issues that relate to
the consequences of the formation of such halos; we proceed to
address them next.

\section{Discussion}

We have derived halo mass fractions in NSs and BHs for the
scenario described in \S2, and the baryonic requirement for
dilution of the ejecta to solar abundances for two cases of very
metal-poor stars. As this population of compact objects is
created astrophysically, this scenario is subject to several
observational constraints, including Big Bang nucleosynthesis
(BBN) limits on the baryon density, and background light
limits. Also of relevance are measurements of the mass and
metallicity of gas presently in the Galaxy, the LG and other
systems, and the role played by subsequent stars in enriching
the Galaxy.

First, as shown in the previous section, for the ejecta from
stars that formed M31 and the Galaxy to not overenrich the
ambient gas with respect to solar abundances, we need a large
amount of primordial gas that was not processed through stars
that could dilute the enriched ejecta. For both $Z = 10^{-4}
Z_\odot$ and $Z = 0$ stars, that translated into an initial
baryonic reservoir of $\sim$ $2-8 \times 10^{12}$ M$_\odot$.
BBN constrains the cosmological density of baryons $\Omega_b
\equiv {\rho_b \over \rho_c}$ to lie in the range $ 0.01 \leq
\Omega_b h^2 \leq 0.024$, where $h \equiv H_0$ /(100 km/sec/Mpc)
(Copi et al. 1995).  Here $\rho_c$ is the critical density
required to close the Universe, $\rho_c = 2 \ h^2 \; 10^{-29}$ g
cm$^{-3} \simeq 3 \ h^2 \times 10^{11}$ M$_\odot$ Mpc$^{-3}$;
for a range of 0.4 to 1.0 for $h$, $\Omega_b$ can vary from 1\%
to 15\%. Recent measurements of the primordial deuterium
abundance in high-redshift quasar absorption systems have
yielded $\Omega_b$ $\simeq$ 0.02 $h^{-2}$ (Burles \& Tytler
1998), which gives a density in baryons of $\rho_b \simeq 6
\times 10^9$ M$_\odot$ Mpc$^{-3}$.

Given $\rho_b$, and that the LG is required to have assembled
from the range of initial baryonic masses in Table 2, we see
that a baryon-gathering radius of $\sim$ 10 Mpc is
necessary. This volume contains a baryonic mass of about 3
$\times 10^{13}$ M$_\odot$, while the dynamical (i.e. total)
masses of all the groups of galaxies present in this volume,
including the LG, is estimated at about 8 $\times$ 10$^{13}$
M$_\odot$ (Tully 1987). We assume that there is no relative
baryon concentration in the LG, i.e., the LG will receive some
fraction of the baryons in the 10 Mpc-radius volume, depending
on the estimates for the LG's total mass. Such a partitioning
will not be in conflict with the existence of field galaxies or
groups around us, and it can then be compared to the
requirements for dilution from Table 2.

One measurement of the dynamical mass of the LG, based on the
least action principle (LAP), gives $M_{\rm LG} \sim$ 6 $\times
10^{12}$ M$_\odot$ (see, e.g., Peebles 1993). This is 8\% of the
dynamical mass in galaxies in the volume of radius 10 Mpc and
gives the LG about 2 $\times 10^{12}$ M$_\odot$ in baryons,
which is close to the lowest mass in Table 2. In this case, only
the $Z = 0$ progenitors which have $Z_{\rm gas,f} = Z_\odot$ are
permitted, and even then just barely. In another analysis
involving the LAP, Dunn \& Laflamme (1993) find that $M_{\rm LG}
\sim 2.2 \times 10^{13}$ M$_\odot$. For this mass, the LG has
28\% of the dynamical mass in the galaxies in a volume of radius
10 Mpc, and thus receives just over 8 $\times 10^{12}$ M$_\odot$
in baryons. This number is approximately equal to the highest
mass in Table 2 and roughly all the cases in Table 2 are
permitted.  Therefore, if we require the LG to have assembled
from the range of initial baryonic masses in Table 2, at least
two cases are not in conflict with the existence of field
galaxies or groups in a volume of radius up to 10 Mpc.

There may be a large amount of unprocessed gas (mostly hydrogen)
and/or processed gas from early phases of star formation in the
intragroup medium today. Depending on how much primordial gas
mixed with the ejecta of early stars or became locked in later
stellar generations, the present gas content of the LG could be
about a few times 10$^{11}-10^{12}$ M$_\odot$ (Table 2). Gas
remaining from forming the Group may not have been completely
retained if the gravitational potential of the LG is
insufficient. If this gas is still present, it must be cooler
than current experimental limits on soft X-ray background
detections. Unfortunately, the total mass of hydrogen in the LG
is not very well-known.  From 21 cm observations, the Galaxy is
constrained to have about 5 $\times$ $10^9$ M$_\odot$ in neutral
hydrogen (Henderson, Jackson \& Kerr 1982).  However, recent
observations of high-velocity clouds (HVCs) which contain
neutral hydrogen indicate that these clouds alone, assuming that
they are gravitationally bound, may have a total mass of $\sim
10^{10}$--$10^{11}$ M$_\odot$, depending on the amount of dark
matter that exists in the Galaxy (Blitz 1997).  The fraction of
the intragroup medium represented in the HVCs is more uncertain,
but it is possible that these clouds are all that is left today
of the initial gaseous reservoir that formed the LG.

In this work, we assume that an early phase of star formation is
widespread, with efficient mixing occuring outside the
star-forming regions, ensured by the effects of heating and
winds from SNe. Observations of present starburst galaxies do
show that they can drive large-scale winds which do not
necessarily diminish the ongoing massive star formation
(Heckman, Armus \& Miley 1987). Therefore, for the LG, we expect
any leftover gas, {\it if } it has remained trapped in the Group
potential, to be distributed in the intragroup medium, and to
have cooled to $10^5-10^6$ K. The upper limit is set by
constraints from diffuse soft X-ray emission and the lower limit
by the absorption features expected from any large amount of
neutral hydrogen in the LG, which are not seen.  In this
temperature range, the intragroup medium is least constrained at
10$^6$ K by specific emission features such as O VI, as compared
to $\sim 10^5$ K where the hydrogen column density is limited to
$\sim$ 10$^{17}$ cm$^{-2}$, or a mass of less than 10$^9$
M$_\odot$ over an area of 1 Mpc$^2$ for the LG.  The presence of
large amounts of intragroup gas at temperatures of about 10$^6$
K and with $Z \ga 0.3 \; Z_\odot$ is therefore an important
target for future observations.

We now consider contributions to background light from the
progenitor stars. The main-sequence lifetime of a star varies
with its mass as: $ \tau_{\rm MS} \sim 10^{10}$
$(M_*/M_\odot)^{-2.2}$ yr (see, e.g., Mihalas \& Binney
1981). All the stars in this scenario have $M \geq$ 10
M$_\odot$, or $\tau_{\rm MS} \la$ 6 $\times$ 10$^7$ years. The
reduced opacity in low-$Z$ stars leads to higher luminosities
which may shorten the lifetime estimate even further (Adams \&
Laughlin 1997). We therefore expect none of these stars to be
burning in the halo today. However, the luminosity from these
massive stars and their deaths as SNe was very high in the
past. To estimate the average luminosity from this bright phase,
we can assume that the LG processed 2 $\times$ 10$^{12}$
M$_\odot$ into 25 M$_\odot$ stars, the median stellar mass in
our IMF. Since the stellar lifetimes for the scenario depicted
here are all less than a few tens of millions of years, this
epoch probably did not last for more than $\sim$ 1 Gyr. The
average luminosity was then about 8 $\times \; 10^{46}$
erg/sec. The surface brightness associated with this luminosity
over a protogalactic region of radius 1 Mpc, is 2 $\times
10^{-4}$ erg/sec/cm$^2$/sr. For the SNe, which are created from
stars up to 60 M$_\odot$, the surface brightness produced, given
the above assumptions, with $L_{\rm SN} \sim 3 \times 10^{42}$
erg/sec for 100 days, is 2 $\times 10^{-7}$
erg/sec/cm$^2$/sr. The SNe are so short-lived that a
time-averaged luminosity is difficult to assign; however,
because of the large amount of gas and dust present in the early
stages of galaxy formation, conditions may be optically thick
enough to lend credence to reprocessed fluxes that are
relatively steady.

To compare these surface brightnesses to current limits on
background light, we assume that the luminosity from the stars
and the SNe are emitted over a wavelength range of 0.05--0.5
$\mu$m. This light will be redshifted, and for a source at
redshift $z$, the net surface brightness (over all frequencies)
is dimmed by a factor of $(1 + z)^4$ (see, e.g., Peebles 1993).
For light reprocessed into infrared wavelengths, we use the
current limits from the FIRAS and DIRBE instruments from the
COBE (Cosmic Background Explorer) satellite. FIRAS (Far Infrared
Absolute Spectrometer) places a limit of 3.4 ($\lambda/400 \mu
m)^{-3}$ nW/m$^2$/sr in the 400--1000 $\mu$m range (Puget et
al. 1996), or an average far infrared background of about
10$^{-6}$ erg/sec/cm$^2$/sr. From Figure 3 of Kashlinsky, Mather
\& Odenwald (1996; references therein), a conservative value for
the infrared background measured by DIRBE (Diffuse Infrared
Background Experiment), after foreground subtraction, is about
10 nW/m$^2$/sr $\simeq 10^{-5}$ erg/sec/cm$^2$/sr, over
wavelengths of 1.25--100 $\mu$m. Light emitted by a source in
the 0.05-0.5 $\mu$m range would be redshifted into the DIRBE
range for $z \ga 1$; to be detected by FIRAS, such a source
would have to be at extremely high redshifts, or have the
stellar light undergo dust reprocessing. Comparing the surface
brightness from the stellar and SNe contributions derived above
with these backgrounds, we see that the light from the SNe is
well below both the FIRAS and DIRBE limits, even before
including the $(1 + z)^4$ dimming. For the progenitor stars,
sources at redshift $\ga$ 1 (3) would be compatible with the
DIRBE (FIRAS) backgrounds. This assumes that the light comes
directly to us, dimmed by $(1 + z)^4$, and that dust, if
present, absorbs and reradiates 100\% of the source's emission;
this is a reasonable upper limit to the contributions of an
early epoch of intense star formation dominated by massive stars
and the resulting SNe. Though the infrared backgrounds constrain
any period of initial massive star formation to satisfy $z \geq
$ 1 or 3, the onset of this phase could follow recombination or
parallel that of galaxy formation.

We had mentioned earlier that the halo mass fraction in NSs and
BHs in the Galaxy (Table 1), $f$, should be viewed as an upper
limit as all such remnants are assumed to ultimately reside in
the halos of the Galaxy and M31. One way this could happen would
be if massive stars were formed in small protogalactic clumps at
high redshifts and blew out their enriched ejecta through SN
winds as galaxies were assembling. The clumps would later merge,
leading to present halos that contain the remnants but not the
gas leftover from making them.  This sequence is similar to that
described in Fields et al. (1997), where the bulk of the
elements were formed at $z \sim 10$.  Alternately, this
starburst phase could occur in galaxies undergoing monolithic
collapse starting at $z = 3-4$; the picture of an evaporating
enriched hot gas component created by SNe, with remnants trapped
in the largest galactic halo potentials, is still valid in the
context of structure formation. In this event, dust in the host
systems must be invoked as being responsible for significant
extinction of this superluminous phase (see, e.g., Mushotzky et
al. 1996, and references therein, for these points). After the
hot ejecta have been blown out of small protogalactic clumps, or
the massive protogalaxies, the gas that remains as a cold gas
component could then either merge, or collapse, to form the
disks of present-day spiral galaxies. The self-enrichment of
this part of the Galaxy is a separate process from the earlier
epoch of Type II enrichment that we consider.

In order to maximize the halo population of NSs and LMBHs, we
focused on an early generation of massive stars. We assumed that
a significant fraction of the baryons were processed through
this early phase; however, it is clear that this scenario does
not account for every baryon in our Galaxy. For instance, a
small fraction of the total baryonic mass needs to form the disk
of our Galaxy where low- and intermediate-mass stars play a
crucial role.  Observations of the present interstellar medium
(ISM) in galaxies help quantify the role played by such stars in
local chemical enrichment. For example, the relation between $Y$
and $Z$ satisfies $\Delta(Y)/\Delta(Z) \sim$ 4 $\pm$ 1 (Timmes
et al. 1996). From Table 2, we see that this quantity, averaged
over the two values of the primordial $Y$, is about 5 for case 1
and 1.7 for case 2. Case 1 is consistent with the observed
value, but case 2 has a somewhat low value of
$\Delta(Y)/\Delta(Z)$.  Case 2 is similar to what we would
expect {\it a priori} of the yields purely from massive
stars. In general, stars of mass less than 8 M$_\odot$ produce
the bulk of the helium from stellar nucleosynthesis, while stars
more massive than this produce the bulk of the metals. The
critical issues in connecting the ISM abundances with those in
an intragroup or intergalactic medium are the assumptions made
for the star formation history, and for how much early
enrichment is reflected in the stars or ISM of present galaxies
due to the role played by outflows and winds. Since the stars
associated with late enrichment are a small fraction of the
total baryons, the gist of our conclusions should not be
altered.

Other baryons we do not account for in this work are those
responsible for metal-poor low-mass stars in the Galactic
halo. The total mass in metal-deficient Galactic halo stars
today is less than about 10$^{10}$ M$_\odot$, which is a small
fraction of the material processed through the early generations
of massive stars discussed here. The primordial gas reserves are
not significantly decreased in forming such metal-poor stars and
those having low masses have not yet returned any material to
the ISM.  An understanding of the physical conditions that
produced the two apparently distinct stellar populations in the
Galactic disk and halo remains to be achieved.

\section{Conclusions}

We have shown that it is possible for halos with significant
mass fractions in NSs and BHs to form in the LG without
necesarrily overenriching the ambient gas relative to solar
$^4$He and/or metal abundances, if these objects were produced
by an early generation of stars with initial metallicity varying
between $10^{-4} Z_\odot$ and 0. In such cases, these remnants
contribute at most 29\% to 36\% of the Galactic halo's dynamical
mass (assumed here to be 10$^{12}$ M$_\odot$). We find, within
the uncertainties in stellar structure and evolution, that
models for halo NSs and BHs cannot be immediately rejected on
metal overproduction arguments alone, due to the critical factor
of the choice of progenitor metallicity in determining the
yields. The constraint of final $Z$-values of $Z_\odot$ for the
ambient gas after dilution was motivated by the metallicities
$Z_X$ observed in the diffuse X-ray gas in galaxy clusters and
groups, which are in the rough ballpark of solar. The baryonic
requirements in Table 2 will scale as $Z_X^{-1}$, if the
dilution criterion is adjusted to other values in the range of
observed X-ray gas metallicities. Furthermore, though $Y$ is not
measured in this X-ray gas and is hence not a direct constraint,
we found that $^4$He is {\it more} constraining than metals as
far as dilution requirements are concerned for low-$Z$ massive
stars' ejecta (Table 2).  The more conventional picture of the
extreme metal pollution from massive stars is true only for
higher metallicity progenitors, such as $Z = Z_\odot$ stars
(Table 1).

NSs make up at least 24\% of the total mass fraction of these
compact objects in the Galactic halo.  LMBHs may be formed from
stars as massive as 40 M$_\odot$ (WW95), as shown in Figures 1
and 2, depending on the kinetic energy chosen for launching the
SN shock in high-mass stars, though we took the most
conservative value for this.

The fraction of NSs and BHs in the Galaxy can be as high as the
values in Table 1, by satisfying the requirement that the LG
attains solar metallicity on average; a preferential loss of
high-metallicity gas to the intergalactic medium need not be
invoked. The most favorable case of this ($Z_{\rm gas,f} =
Z_\odot$) is for $Z=0$ stars where $f$ = 36\% (Table 1), and
$M_{\rm gas,i} = 2.4 \times 10^{12}$ M$_{\odot}$ (Table 2). The
worst case is that of the 10$^{-4} \, Z_\odot$ stars with
$Y_{\rm gas,f} = Y_\odot$, where $f$ = 29\% and $M_{\rm gas,i} =
8.1 \times 10^{12}$ M$_{\odot}$. We emphasize that this derived
$f$ should be viewed as an upper limit, as we have assumed that
all the remnants created in this scenario ultimately reside in
galaxy halos till today. This assumption is plausible in the
context of current structure formation scenarios (\S4). However,
if we were to be more democratic, and preserve, on the scale of
galactic halos, the global baryon-to-{\it total} mass ratio,
then the Galaxy and M31 would each receive a fraction ($M_{\rm
gal}/M_{\rm LG}$)$\ast f$ in remnants. Then, $f$ would be
reduced in the ``best'' and ``worst'' cases above to about 6\%
and 1\% respectively.

$Z=0$ progenitor stars have lower integrated metal yields
and leave behind higher masses in remnants (Table 1), but as
some self-enrichment is expected to occur in early star-forming
regions, the case of $Z = 10^{-4} Z_\odot$ stars is probably
more realistic. This however needs larger baryonic masses for
dilution to avoid supersolar pollution, which would require a
high adopted mass for the LG as derived by Dunn \& Laflamme
(1993).  Nevertheless, it is possible, though just barely, to
reconcile the baryonic requirements set by Table 2 with BBN
constraints, without assuming a relative baryon concentration in
the LG via cooling processes, etc. This is especially true for
dilution of metal enrichment to solar levels. Other limits such
as background light production are met more comfortably as shown
in \S\ 4. As we have discussed, the presence of massive stars at
early epochs are favorable from some aspects, but no one stellar
population can explain all the observations.

We have focused on an initial burst involving only high-mass
stars in order to maximize the contribution of NSs and LMBHs to
halo MACHOs, and calculated the yields from just this epoch. In
doing this, we have assumed that the baryons that form Pop. I
and II stars are not enriched by the burst which creates the
MACHOS in this scenario and leaves behind a large amount of
enriched gas. This segregation is possible if galactic winds
which accompany the burst phase remove associated polluted
ejecta, leaving a small fraction of material behind that
eventually forms disk and halo stars. The combined Pop. I/II
baryonic mass is at most a few percent of that associated with
the burst phase, and will not alter the main conclusions of this
paper. A similar separation in enrichment may have occured in
the case of the hot X-ray gas, seen in many galaxy clusters and
groups, which contains a significant fraction, and often most,
of the baryons. This gas is relatively enriched in metals and
co-exists with galaxies containing low-$Z$ stars.

Though some recent observations have made a re-evaluation of
baryonic dark matter candidates necessary, an unavoidable
consequence of making a case for stellar remnants as MACHOs is
that they are only a fraction of the mass processed through
their parent stars, and a significant amount of enriched gas
will remain. Adams \& Laughlin (1996) have noted this efficiency
problem in creating halo WDs, and this is especially true for
NSs/BHs relative to WDs. The early generations of stars leave
behind significant amounts of enriched ejecta. It is possible
that much of this enriched gas escaped altogether from the Group
or was ejected into intergalactic space (as suggested in Adams
\& Laughlin 1996, and Fields et al. 1997). Here, we have shown
that if we want to dilute $Y$ or $Z$ in the leftover gas to
solar values, then the LG most likely started from a baryonic
reservoir $\ga$ 2 $\times$ 10$^{12}$ M$_\odot$ (\S\ 3).  We have
made a plausibility case for NS/BH halos that do not leave
behind overenriched gas. However, the present uncertainties in
supernova yields, the mass of the LG, and the primordial $^4$He
abundance could affect our results. The contribution of NS/BH
remnants to galaxy halos may be smaller than those in Table 1,
or the leftover intragroup gas could have lower
metallicities. For progenitor metallicities $\la 10^{-4}
Z_\odot$, the total masses in the remnant population in our
results are in the range of the experimental most likely total
mass in MACHOs of $\sim$ 2 $\times$ 10$^{11}$ M$_\odot$,
mentioned in \S\ 2.

It is encouraging that the metal abundance of the gas remaining
after this phase of early star formation is $Z \ga $ 0.3
$Z_\odot$, which is not in conflict with that of the intragroup
gas in some groups of galaxies (Mulchaey et al. 1996). After the
Group formed, a large amount of gas is left over in this
scenario. If the potential of the LG can retain this gas, it
must exist in a diffuse component in the intragroup medium at
temperatures $\sim 10^6$ K; the Group's potential is probably
insufficient to heat it to keV temperatures. This is consistent
with the idea put forth by Mulchaey et al. (1996) that the
intragroup medium in spiral-rich groups could have $T \sim$ 0.2
keV, and remain invisible to current experiments such as ROSAT.
Studies of clusters of galaxies show that there is up to several
times as much mass in just the X-ray emitting gas as the mass in
the galaxies (see e.g., Henry, Briel \& Nulsen 1993). It remains
to be seen how groups of galaxies, especially spiral-dominated
ones, compare in this regard, especially if the potential of the
group is not large enough to keep any gas that is present heated
at keV temperatures (Mulchaey et al. 1996).

Alternatively, some of the leftover gas could be locked into
faint low-mass field stars throughout the intragroup medium, or
some gas may have been lost from the Group. In the latter case,
the gas must remain ionized; otherwise, it would violate
Gunn-Peterson constraints on the amount of intergalactic neutral
hydrogen. The recent detections of $\sim 10^6$ K gas by the
Extreme-Ultraviolet Explorer in the directions of the Virgo and
Coma clusters might prove to be important in resolving the issue
of the fate of such gas in nearby systems. In the case of Virgo
(Lieu et al. 1996), this gas appears to be relatively enriched
($\sim 0.5 \; Z_\odot$). Another key issue is whether the
elemental abundance ratios in the hot gas in clusters of
galaxies indicate a Type II or Ia SN origin; at present, it is
undecided. If a Type II SN origin for this gas emerges in future
studies, it would certainly favor the presence of a remnant
population dominated by NSs and BHs. We conclude therefore that
most of the baryonic ``dark matter'' might be in cool ($\sim
10^6$ K) gas in the LG, but we emphasize that stellar remnants
in the form of MACHOs are also a significant factor in the
galactic dark matter problem.

Halo BHs are not expected to be significant sources of
radiation, for typical halo ambient gas densities. However, BHs
from this population have a local number density of $\sim$ 2
$\times 10^{-5}$ pc$^{-3}$, assuming that they are distributed
in a 100 kpc halo today. Therefore, there could be at least one
such BH within a 50 pc radius from us. If it has sufficiently
low velocity, it might be detected in future projects such as
the Sloan Digital Sky Survey (Heckler \& Kolb 1996).

As we have seen, evaluating the case for stellar remnants as
MACHOs is not free of some assumptions. The IMF from which the
MACHOs were created must be different from that of present
galactic disks. The IMF's peak must be tailored to $\sim$ 2--4
M$_\odot$ for WDs, or to include only stars more massive than
about 10 M$_\odot$ for NSs and BHs. Furthermore, all candidates
face some critical issues. To not overenrich the Universe, some
dilution of metal-rich SN ejecta must take place, and this
constraint is considerably tighter for NS/BH remnants than for
WDs, simply due to the fraction of the progenitors that
collapses into the remnant in each case. If dilution does not
occur, and the enriched gas is ejected from galaxies into an
intracluster or intergalactic medium, strong constraints, from
measured iron and oxygen abundances in the intracluster medium
or from the carbon enrichment in the Lyman-alpha forest clouds
(particularly for WD models), must be considered. (See Fields,
Freese \& Graff (1998) for a comprehensive discussion on the
relative merits and limitations of these baryonic MACHO
candidates.) The burst of intense star formation may be
unobservable if it occurred at sufficiently high redshifts or if
it is obscured by dust.  However, halo WDs must contend with
limits from the Hubble Deep Field, which constrains their age
via their cooling emission, and, with greater model dependency,
the current measured rates of Type Ia SNe, while halo NSs and
BHs contribute a direct signature primarily through their mass.
 
The MACHO experiment's results indicate that about 50\% of our
halo may be made of objects of mass $\sim$ 0.5 M$_{\odot}$ (for
a ``standard halo'' mass of 4 $\times$ 10$^{11}$ M$_\odot$
within 50 kpc).  Experiments with longer periods of observation
may detect events of longer duration. The typical MACHO mass is
unlikely to be lower than the presently quoted 0.5 M$_\odot$, as
strong constraints placed by the MACHO and the EROS
collaborations (Renault et al. 1997; Alcock et al 1998), based
on shorter timescale events, have already ruled out significant
populations in objects of lower masses.  Also, Graff \& Freese
(1996a, 1996b) have ruled out significant halo mass fractions in
red dwarfs and brown dwarfs. Gyuk \& Gates (1998) have shown
that even models of rotating halos cannot lower the present
experimental MACHO mass to the brown dwarf range. An increase in
the average MACHO mass is however possible if more events of
longer duration are detected with time. In fact, the MACHO
candidate mass has increased over the last few years (Alcock et
al. 1993, 1996, 1997). Therefore we have considered the
possibility of MACHOs in the halo of $\sim$ solar and somewhat
higher masses. 

If such objects exist, they may be seen in the future by
microlensing experiments as longer-timescale events. In order to
have some idea of the microlensing signature of halo NSs and
BHs, we derive here some rough numbers for the MACHO experiment
in particular. In this work, roughly 85\% of the NS/BH remnants
produced have masses $\la$ 5 M$_\odot$. The duration of a
microlensing event is mass-dependent; for a standard halo model,
the average event timescale varies with the MACHO mass as,
$<\hat t>$ $\propto$ $\sqrt{M D}/v \ \cong 140 \sqrt{M/M_\odot}$
days (Alcock et al. 1997), for an assumed MACHO velocity $v$ and
distance $D$. Thus NSs/BHs of mass $\la$ 5 M$_\odot$, if they
exist in the Galactic halo, will be picked up by the MACHO
experiment as events with average timescales $\la$ 313 days
$\sim$ 0.86 yr. A typical NS, or a LMBH, of mass $\sim$ 1.5
  M$_\odot$, would have $<\hat t>$ $\sim$ 150 days. However, the
most {\it probable} duration of events from 1.5 M$_\odot$
objects in the halo is only $\sim$ 90 days (Griest 1991). The
MACHO experiment, subsequent to the 2-year data publication, has
detected a few events with durations approximately between 90
and 105 days. While this is hardly a proof of the existence of
halo NSs and BHs, it is worth noting in relation to the question
of how such objects would appear to microlensing experiments,
and whether they may have already been detected. Given our
prediction of longer timescale events from halo NSs and BHs, it
is provocative that the first-year data of EROS2 in the
direction of the SMC give a most probable lens mass of over 2
M$_\odot$ if they are present in the Galactic halo
(Palanque-Delabrouille et al. 1998), though self-lensing by the
SMC is also a possibility. Observations of the subsequent event
SMC-98-1, however, seem to imply a binary lens located in the
SMC itself (see, e.g., Afonso et al. 1998).

As the final halo mass fraction in NSs/BHs in this work is less
than 100\%, it is not necessarily incompatible with the presence
of other components which could account for the shorter-duration
microlensing events that are already detected. Obviously, one
can only take the theory of two or more distinct halo
populations to a point, beyond which the measured dynamical halo
mass is exceeded. We have seen above an estimate of the
microlensing signature from halo NSs and BHs. An alternative
question is how the present microlensing data can constrain any
higher-mass population. In the context of the standard halo
model, the data indicate a most likely lens mass of about 0.5
M$_\odot$. However, the lens mass estimate depends upon the
phase space distribution of the lenses within the halo. To
constrain any NS-mass population, we must explore the parameter
space for galactic halo models, given the current data; we
consider this problem in a subsequent work.

\acknowledgements

We gratefully acknowledge illuminating discussions with
B. Holden, M. Lemoine, and F. X. Timmes, and our referee
B. Fields for his many useful suggestions. We also thank
E. Gates, K. Griest, A. Meiksin, M. C. Miller, and D. G. York
for helpful comments. A. V. and A. V. O. were supported in part
by the DOE through grant DE-FG0291 ER40606, and by the NSF
through grant AST-94-20759. A. V. acknowledges the support of
the Farr Fellowship at the University of Chicago.

\newpage
\begin{figure}
\plotone{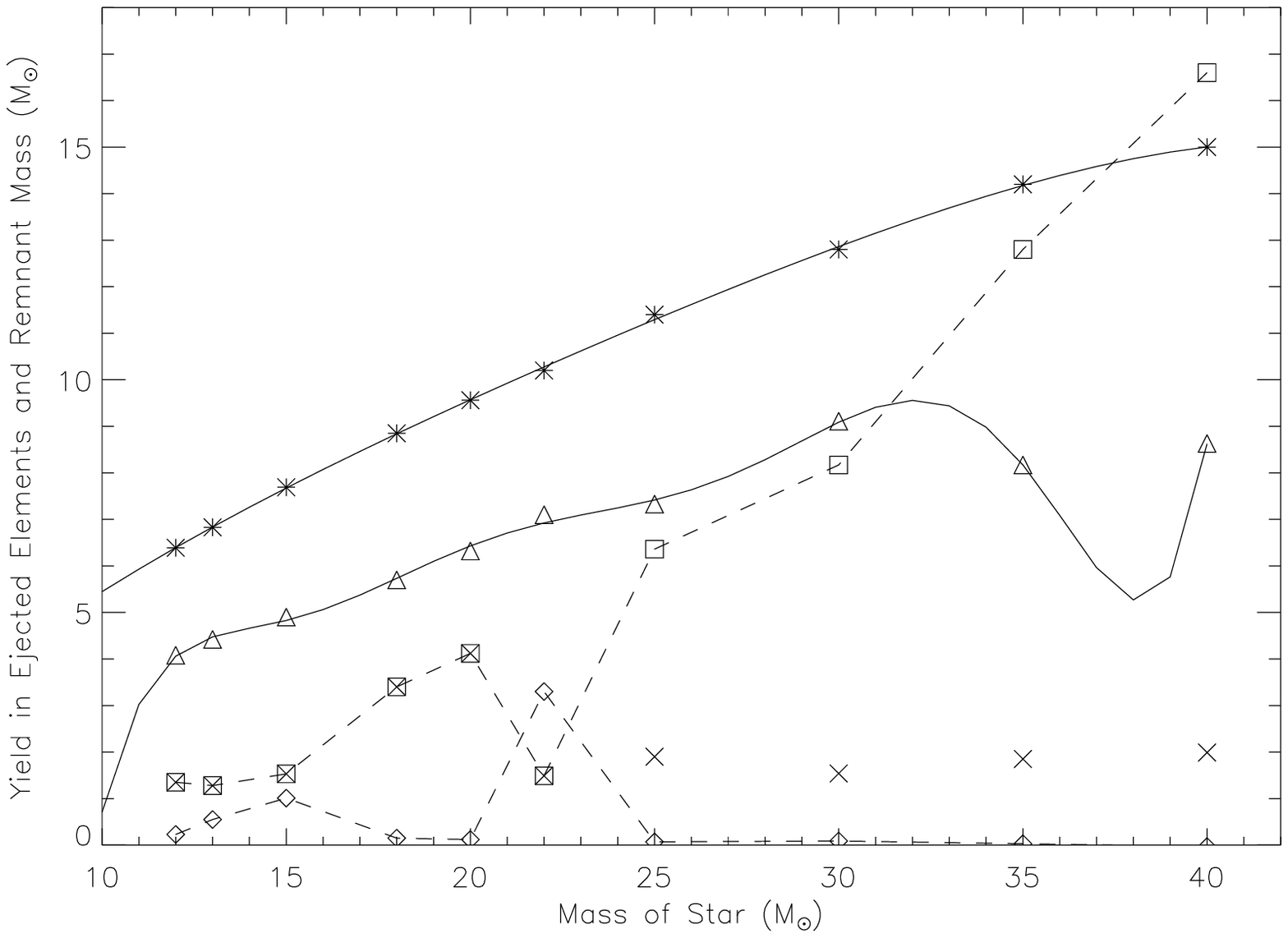}
\caption{{\it For $Z = 0$ stars from WW95}: Ejected
masses in hydrogen (plotted as asterisks), $^4$He (triangles),
and metals (diamonds). Also shown are the masses in
post-supernova remnants for the minimal and maximal ejecta
models (squares and crosses respectively) from WW95; the dashed
lines are drawn through to guide the eye for the diamonds and
squares. The solid lines, where drawn, represent the polynomial
fit for that case. All quantities are in solar mass units.}
\end{figure}

\begin{figure}
\plotone{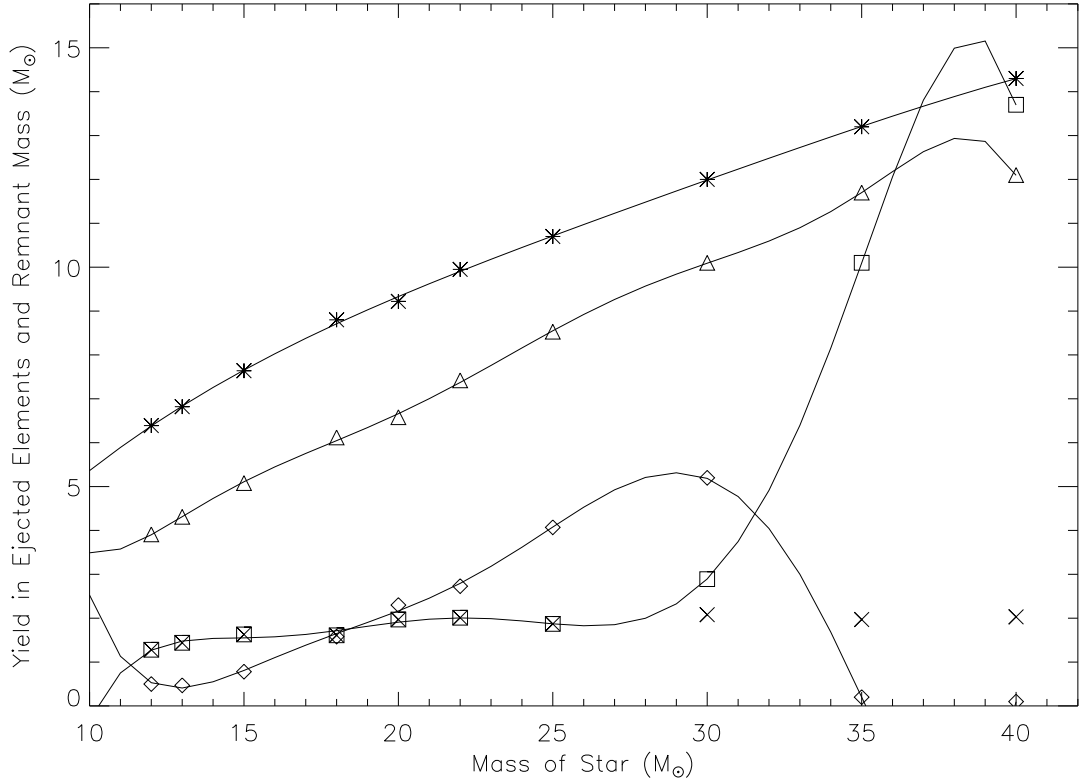}
\caption{{\it For $Z = 10^{-4}\; Z_\odot$ stars from
WW95}: Ejected masses in hydrogen (plotted as asterisks), $^4$He
(triangles), and metals (diamonds). Also shown are the masses in
post-supernova remnants for the minimal and maximal ejecta
models (squares and crosses respectively) from WW95. The solid
lines represent the polynomial fit for each case. All quantities
are in solar mass units.}
\end{figure}

\clearpage

\begin{table}
\begin{center}
\begin{tabular}{||c|c|c|c|c|c|c|c|c||} \hline
 & $Z_{\rm star}$ & $M_{\rm H}$ ($M_\odot$) & $M_{\rm ^4He}$ ($M_\odot$) &
$M_{\rm Z}$ ($M_\odot$) & $M_{\rm rem}$ ($M_\odot$) & $f$ & $Y_{\rm
gas}/Y_\odot$ & $Z_{\rm gas}/Z_\odot$ \\  \hline
Case 1 & 0  & 3.9 $\times 10^{11}$ & 2.35 $\times 10^{11}$ & 1.59
 $\times 10^{10}$ & 3.6 $\times 10^{11}$ & 36\% & 1.33 & 1.31
 \\ \hline
Case 2 & $10^{-4}$ $Z_\odot$ & 3.78 $\times 10^{11}$  & 2.77 $\times
10^{11}$ & 6.32 $\times 10^{10}$ & 2.89 $\times 10^{11}$ & 29\% & 1.40
& 4.65 \\ \hline
Case 3 & $Z_\odot$ (mE) & 3.35 $\times 10^{11}$ & 2.86 $\times 10^{11}$ & 9.76
$\times 10^{10}$ & 2.81 $\times 10^{11}$ & 28\%  & 1.45 & 7.18 \\ \cline{3-9}
 & $Z_\odot$ (ME) & 3.35 $\times 10^{11}$ & 2.86 $\times 10^{11}$ & 1.43
$\times 10^{11}$ & 2.36 $\times 10^{11}$ & 24\% & 1.36 & 9.92 \\ \hline 
\end{tabular}
\end{center}
\caption{For three cases of initial metallicity of the
progenitor star, the yields in hydrogen, $^4$He and metals, and
remnant masses in solar mass units, from a stellar population of
mass $10^{12}$ M$_{\odot}$ per galaxy. Also shown are the mass
fractions of the ejected gas in $^4$He and metals relative to
solar values, and the final halo mass fraction $f$ in remnants
for a galactic halo of total mass $10^{12}$ M$_{\odot}$.}
\end{table}

\begin{table}
\begin{center}
\begin{tabular}{||c|c|c|c|c||} \hline 
$Z_{\rm star}$ & $Y_{\rm gas,i}$ & $M_{\rm gas,i}$/$M_\odot$ &
$Y_{\rm gas,f}/Y_\odot$ & $Z_{\rm gas,f}/Z_\odot$ \\  \hline
0 & 0.232 & 4.7 $\times 10^{12}$ & 1 & 0.42 \\ \cline{2-5} 
  & 0.232 & 2.4 $\times 10^{12}$ & 1.22 & 1 \\ \cline{2-5}
  & 0.249 & 6.5 $\times 10^{12}$ & 1 & 0.29 \\ \cline{2-5}
  & 0.249 & 2.4 $\times 10^{12}$ & 1.23 & 1 \\ \hline
$10^{-4}$ $Z_\odot$ & 0.232 & 5.7 $\times 10^{12}$ & 1 & 1.3 \\ \cline{2-5}
  & 0.232 & 7.2 $\times 10^{12}$ & 0.96 & 1 \\ \cline{2-5}
  & 0.249 & 8.1 $\times 10^{12}$ & 1 & 0.89 \\ \cline{2-5}
  & 0.249 & 7.2 $\times 10^{12}$ & 1.01 & 1 \\ \hline
\end{tabular}
\end{center}
\caption{For two cases of initial metallicity of the progenitor
star, and two primordial $^4$He values, the mass of initial gas
required to dilute either $Y$ or $Z$ of the enriched ejecta from
both M31 and the Galaxy to solar values, with the resulting
final values in both $Y$ and $Z$ of the gas also shown, relative
to solar values.}
\end{table}

\clearpage
\end{document}